\documentclass[a4paper]{jpconf}
\usepackage{adjustbox}
\usepackage{graphicx}
\usepackage{tcolorbox}
\begin{document}

\title{The swiss army knife of job submission tools: grid-control}

\author{F Stober$^1$, M Fischer$^2$, P Schleper$^1$, H Stadie$^1$, C Garbers$^1$, J Lange$^1$, N Kovalchuk$^1$}
\address{$^1$ Particle Physics and Detector Development Group, Institute of Experimental Physics, Luruper Chaussee 149, 22761 Hamburg, Germany}
\address{$^2$ Steinbuch Centre for Computing, Karlsruhe Institute of Technology, Hermann-von-Helmholtz-Platz 1, 76344 Eggenstein-Leopoldshafen, Germany}

\ead{fred.stober@cern.ch}

\begin{abstract}
{\it{grid-control}} is a lightweight and highly portable open source submission tool
that supports all common workflows in high energy physics (HEP).
It has been used by a sizeable number of HEP analyses
to process tasks that sometimes consist of up to $100k$ jobs.

{\it{grid-control}} is built around a powerful plugin and configuration system,
that allows users to easily specify all aspects of the desired workflow.
Job submission to a wide range of local or remote batch systems or grid middleware is supported.

Tasks can be conveniently specified through the parameter space that will be processed,
which can consist of any number of variables and data sources with
complex dependencies on each other.
Dataset information is processed through a configurable pipeline of dataset filters,
partition plugins and partition filters. The partition plugins can take the number
of files, size of the work units, metadata or combinations thereof into account.

All changes to the input datasets or variables are propagated through the processing pipeline
and can transparently trigger adjustments to the parameter space and the job submission.

While the core functionality is completely experiment independent,
full integration with the CMS computing environment is provided by a small set of plugins.
\end{abstract}

\section{Introduction}

A common task in high energy physics (HEP) is to run a large number of
scripts or executables on local or remote computing resources. In order to
efficiently process these high throughput computing (HTC) tasks, it is
necessary to automate the submission, monitoring, output retrieval and
smart re-submission of jobs.

{\it{grid-control}} was initially developed in order to facilitate the HTC analysis work
of the CMS group in Karlsruhe on a large variety of available local batch and grid resources
and help with particular maintenance tasks of the Tier-1 data center GridKA.

Tools that were available at the time like CRAB\cite{CRAB} or Ganga\cite{GANGA}
did not support needed functionality like automatic re-submission and CMS dataset queries (for analysis tasks)
or the ability to handle custom job description language attributes and special VOs (for Tier-1 testing purposes).
Subsequently developed job management tools like CRAB3\cite{CRAB3} or Lobster\cite{LOBSTER}
can't submit to some of the local batch systems that are available to users,
enforce particular analysis environments or depend on centralized server infrastructure.

Over the past 10 years, {\it{grid-control}} has evolved from a basic job submission and
monitoring tool into a sophisticated user workflow management suite.
Since most development is done by people directly working on HEP analysis, it
covers practically all aspects of a HEP user analysis.

Supported {\it{grid-control}} applications range from doing multivariate / ntuple analysis with ROOT macros,
submitting scripts to a set of worker nodes or compute elements,
working with official HEP experiment software with multiple generation or processing steps
to running all different kinds of custom software.

Unique features of {\it{grid-control}} are a sophisticated plugin and configuration system,
that allows users to gain complete control of all aspects of the workflow and provide
additional functionality or site / user / task / experiment / analysis specific customisation.

\section{Using grid-control}

\subsection{Requirements and installation}

{\it{grid-control}} can be deployed in virtually any environment with access to some job submission infrastructure.
In order to achieve this, {\it{grid-control}} avoids any external dependencies and restricts itself
to a highly portable subset of Python\cite{PYTHON} language features.
This allows {\it{grid-control}} to run under a wide range of python interpreters like
CPython~2.3~-~3.5 or pypy~2.4~-~5.3.

Job submission to many different kinds of local or remote workflow management software is possible.
Stable support exists for (HT-)Condor\cite{CONDOR}\cite{HTCONDOR}, PBS\cite{PBS}, JMS\cite{JMS},
SLURM\cite{SLURM}, GridEngine\cite{SGE}, glite(-WMS) or CreamCE\cite{CREAMCE}.
There is also some highly experimental support for ARC\cite{ARC} and Dirac\cite{DIRAC} submission.
For small tasks or tests, it is also possible to simply run jobs on the host itself.
Workflows can also be configured to use multiple job submission backends at the same time.

On the worker nodes, only a basic bash environment is needed to run jobs.
However, in order to use HEP experiment software or access grid storage, having access to
gfal tools or CVMFS is suggested.

The most convenient way to install {\it{grid-control}} uses the Python package manager pip:
\begin{verbatim}
pip install --user grid-control
\end{verbatim}
Where the option \texttt{--user} tells the package manager to install it only for the local user.

\subsection{Configuration and execution}

The easiest way to use {\it{grid-control}} is through the command line program
\texttt{gridcontrol} that takes ini-style or python config files.
However for advanced users, it is also possible to directly steer {\it{grid-control}} from a user
script through a simple API.

The following ini-style config file for example instructs {\it{grid-control}} to run the script
\texttt{test.sh} two times in the local batch system:
\begin{tcolorbox}
\begin{verbatim}
[global]
task      = UserTask               ; Job uses user written scripts
backend   = local                  ; Send to local batch system
[jobs]
jobs      = 2                      ; Submit script two times
wall time = 1:00                   ; Jobs will take max 1h
[UserTask]
executable = Example.sh            ; Name of the script
\end{verbatim}
\end{tcolorbox}

The syntax to execute {\it{grid-control}} with this config file is: \texttt{gridcontrol <config file>}.

While {\it{grid-control}} is running, it will check the status of running jobs,
retrieve and analyse finished jobs and submit jobs as needed.
When the continuous mode is activated with \texttt{-c}, this cycle will repeat until the task is finished
or the user manually quits the program.

Further help about possible actions that can be performed on the command line
can be requested with \texttt{gridcontrol --help}.
One example for a commonly used action is to trigger the cancellation of all queued and running jobs
(e.g. because of an issue with the user script). This can be done with the command: \texttt{gridcontrol -d ALL <config file>}.

The command line program is quite often run from within the GNU \texttt{screen} terminal manager,
which allows the user to continuously monitor the status of the running workflow on the console.

\section{Parameterised jobs}

A particularly useful feature for HEP applications is the sophisticated
job parameter system built into {\it{grid-control}}. This system makes it
possible to define the parameter space that a task is based on in a convenient
domain specific language (DSL). For simplicity, only the default DSL is
presented in the following example.

The parameter space composed by the user can be arbitrary complex, combining
any number of user supplied variables and data sources with dependencies
between them.

It is also possible to specify the job requirements (like wall time / memory)
for each parameter space point separately, which allows the optimal utilization
of all available job slots.

An example for such a parameterized workflow is given in the following config file:
\begin{tcolorbox}
\begin{verbatim}
[global] include = HelloWorld.conf ; include file with basic settings
[parameters] ; this section allows to define the main parameter space
parameters = (MUR,MUF) VAR[MUR] + {pspace1}
(MUR, MUF) = (1, 1) (2, 1) (1, 2) ; define tuple with 3 values
VAR = def    ; "def" is the default for this lookup variable
  2 => x y   ; return "x" and "y" if the input variable matches "2"
[pspace1]    ; this section defines a new parameter space called "pspace1"
parameters = MUR MUF ; cross product of MUR and MUF
MUF = 0.5            ; variable with a single value "0.5"
MUR = 1 2            ; variable with two possible values "1" and "2"
\end{verbatim}
\end{tcolorbox}

Running {\it{grid-control}} with this config file gives the
parameter space that is described in table~\ref{PARAM1}.

\begin{table}[ht]
\adjustbox{valign=t}{\begin{minipage}[b]{0.48\linewidth}
\caption{\label{PARAM1}List of parameters for the initial configuration.}
\begin{center}
\lineup
\begin{tabular}{*{5}{l}}
\br
Job & MUR & MUF & VAR & Status \cr
\mr
0 & 1 &   1 & def & active\cr
1 & 2 &   1 &  x  & active\cr
2 & 2 &   1 &  y  & active\cr
3 & 1 &   2 & def & active\cr
4 & 1 & 0.5 &     & active\cr
5 & 2 & 0.5 &     & active\cr
\br
\end{tabular}
\end{center}
\end{minipage}}
\hspace{0.02\linewidth}
\adjustbox{valign=t}{\begin{minipage}[b]{0.48\linewidth}
\caption{\label{PARAM2}List of parameters for the changed configuration.}
\begin{center}
\lineup
\begin{tabular}{*{5}{l}}
\br
Job & MUR & MUF & VAR & Status \cr
\mr
0 & 1 &   1 & def & active\cr
1 & 2 &   1 &  x  & active\cr
2 & 2 &   1 &  y  & active\cr
3 & 1 &   2 & def & active\cr
4 & 1 & 0.5 &     & active\cr
5 & 2 & 0.5 &     & disabled\cr
6 &0.5& 0.5 &     & active\cr
\br
\end{tabular}
\end{center}
\end{minipage}}
\end{table}

In addition, the parameter system is able to handle changes to the data source,
as well as to other parameters, and can transparently adapt the job submission to
the new parameter space at run-time.

This can be demonstrated by changing the last line in the above config file from \texttt{MUR = 1 2} to
\texttt{MUR = 0.5 1} and restarting {\it{grid-control}}.
This will disable (and cancel if needed) job 5 and submit a new job 6.
The new parameter space after this change is shown in table~\ref{PARAM2}.

\section{Dataset processing pipeline}

{\it{grid-control}} provides a wide range of options on how to process datasets.
Several built-in plugins are available to retrieve datasets from different sources
(e.g. text files, (SE) directory listings, databases, REST APIs, other {\it{grid-control}} instances, etc.).

The datasets provided by these sources may contain a list of URLs,
an optional number of work units (e.g. the number of events or file size),
arbitrary metadata (e.g. run / luminosity section information or generator parameters)
and locality information.

\begin{figure}[ht]
\centering
\includegraphics[scale=0.4]{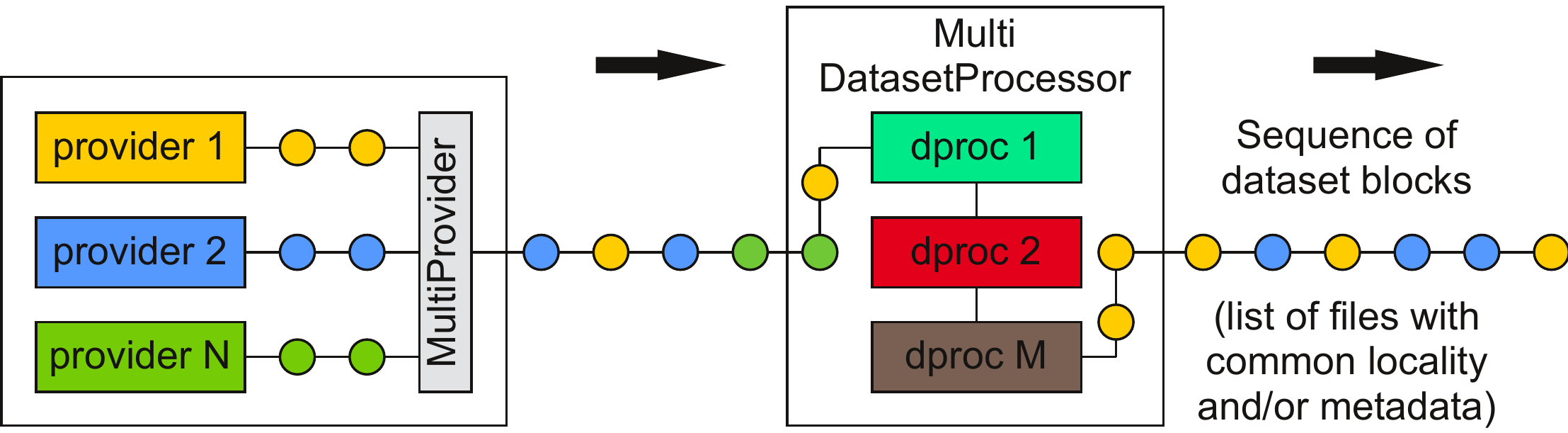}
\begin{minipage}[b]{0.4\linewidth}\caption{\label{DPROV}The first step in the dataset processing pipeline is the
aggregation of results from one or multiple dataset providers in order to form
a stream of dataset information blocks. This stream of blocks is subsequently
filtered or modified by a configurable series of dataset processors.}
\end{minipage}
\end{figure}

\begin{figure}[ht]
\centering
\includegraphics[scale=0.45]{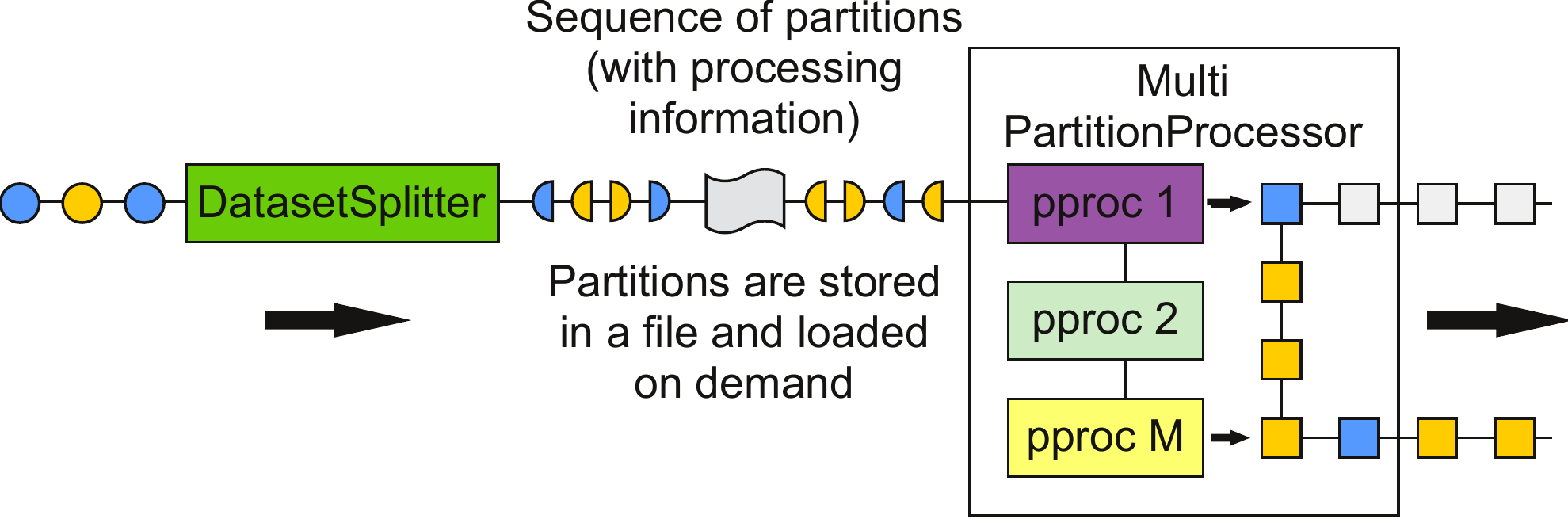}
\begin{minipage}[b]{0.4\linewidth}\caption{\label{PPROC}In the second step,
the prepared dataset blocks are split into partitions. These are subsequently stored in a file and later loaded on demand.
The partition information is later processed by a configurable series of partition processors
which fill the parameter space points with the appropriate information for that particular dataset.}
\end{minipage}
\end{figure}

All datasets are processed through a configurable pipeline of
dataset processors, partition plugins and partition processors as
shown in figure~\ref{DPROV} and \ref{PPROC}.

Dataset processors can for example remove unwanted files matching some particular file name pattern
or metadata information, modify dataset locations according to some blacklist or perform
consistency checks and de-duplications on the input dataset blocks.

\begin{figure}[ht]
\includegraphics[width=0.55\linewidth]{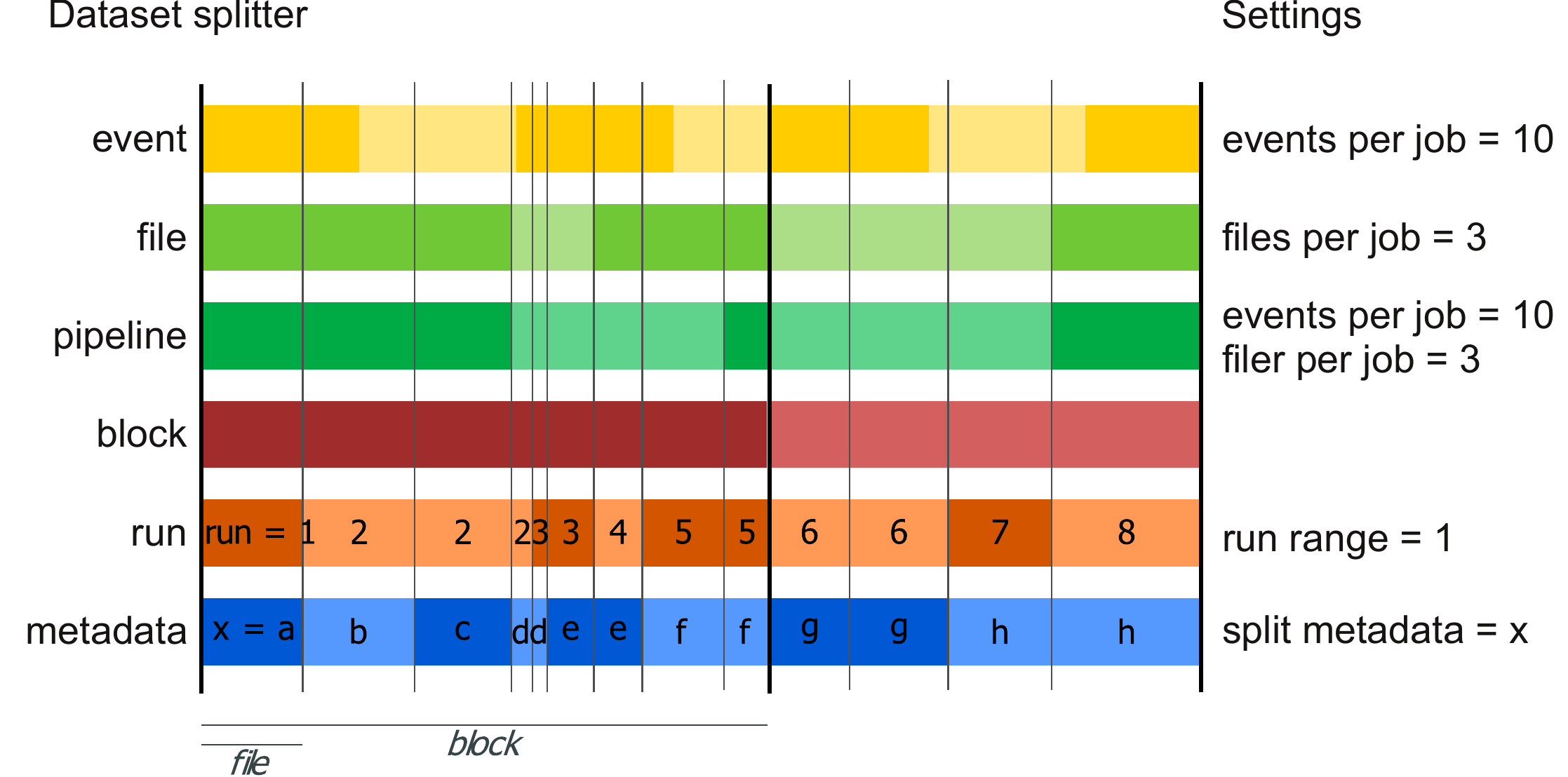}\hspace{0.05\linewidth}%
\begin{minipage}[b]{0.4\linewidth}\caption{\label{PARTITION}Visualization how
different partition plugins can split the stream of dataset blocks into partitions that
form the dataset parameter space.}
\end{minipage}
\end{figure}

In the next step, the processed stream of dataset blocks is given to the partition plugins.
Several methods to split datasets into partitions are supported and the effect of
some of them is shown in figure~\ref{PARTITION}.
These partition plugins can take the number of files,
size of the work units, metadata or combinations thereof into account.

The partition processor translates the partition information
into the variables and requirements of a parameter space point.
It can also perform operations that are similar to the work
done by the dataset processor (e.g. change LFNs into PFNs, apply blacklists to dataset locations, etc.).

Modifications to the original dataset information source, whether it be on file level (e.g. additions or removals)
or on the work unit level (e.g. expanding or shrinking files),
are propagated through the processing pipeline and transparently
trigger adjustments to the processed parameter space.

For HEP workflows this allows to run a "live" analysis on an ever
expanding dataset with a single processing task that regularly
queries some data source and spawns new jobs.

\section{Performance}

In order to evaluate the performance and scaling behaviour of {\it{grid-control}},
the memory consumption (max. resident set size) as well as the user and system
CPU time is measured for different workloads.

The benchmark is performed with a single core on an i5 2.4GHz machine without SSD.
Only the resources used by {\it{grid-control}} itself are measured
(e.g. local job database interactions, preparation of job files, etc.).
This means in particular that the time spent by the underlying job submission
infrastructure (e.g. to transfer files, job registration with the scheduler, etc.)
is not taken into account, since this strongly depends on the used job submission backend.

The workflow is configured to run with 3 different parameters over
a randomly generated data source, which contains 9 datasets, with
each dataset block containing 99 files. For this test, the number of blocks was
varied between 1 and 2000, resulting in up to 300k jobs.

\begin{figure}[ht]
\begin{minipage}{0.48\linewidth}
\includegraphics[width=\linewidth]{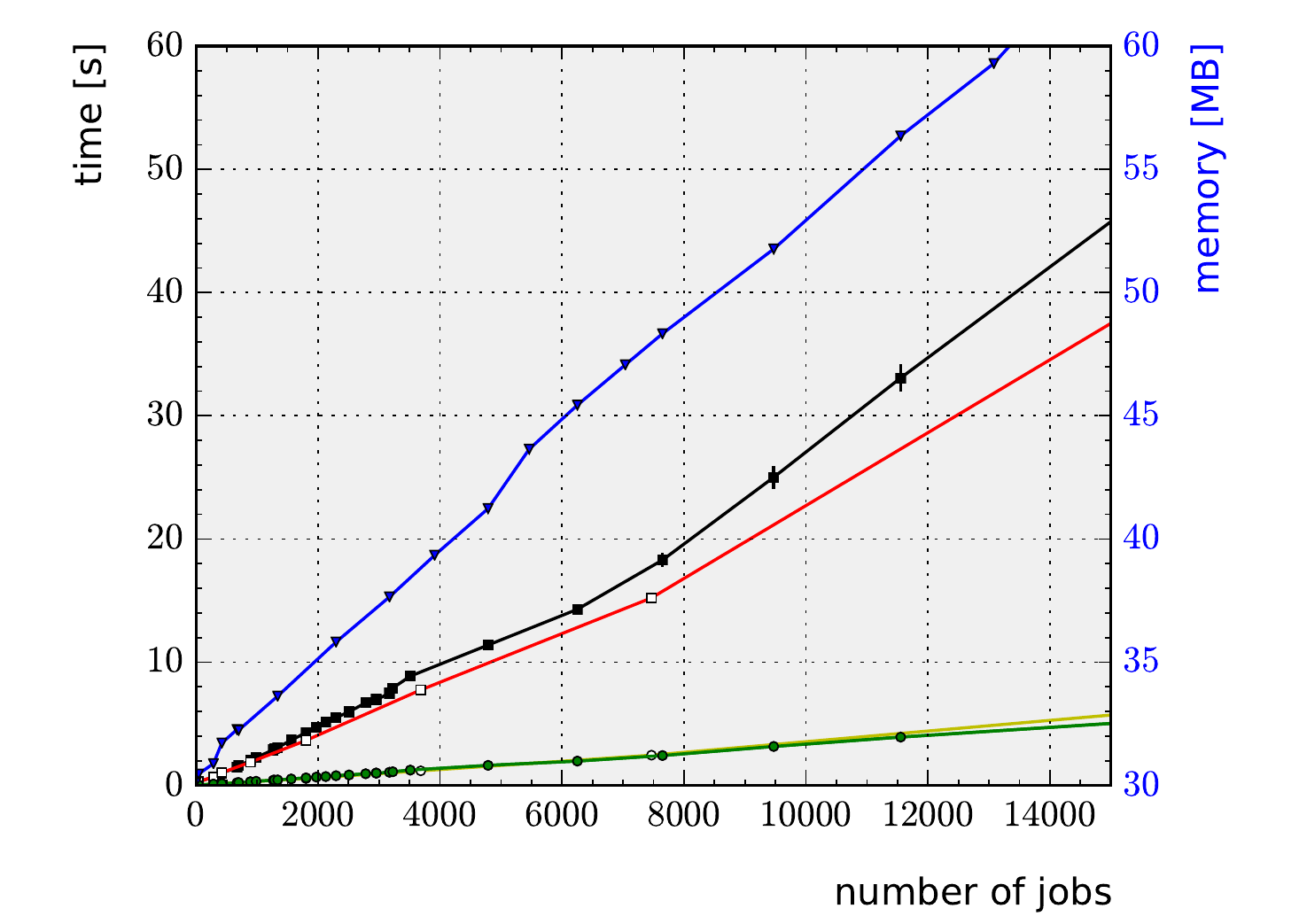}
\caption{\label{SCALING_ZOOM}Memory and CPU time usage for common task sizes of up to 15k jobs.
}
\end{minipage}\hspace{0.04\linewidth}
\begin{minipage}{0.48\linewidth}
\includegraphics[width=\linewidth]{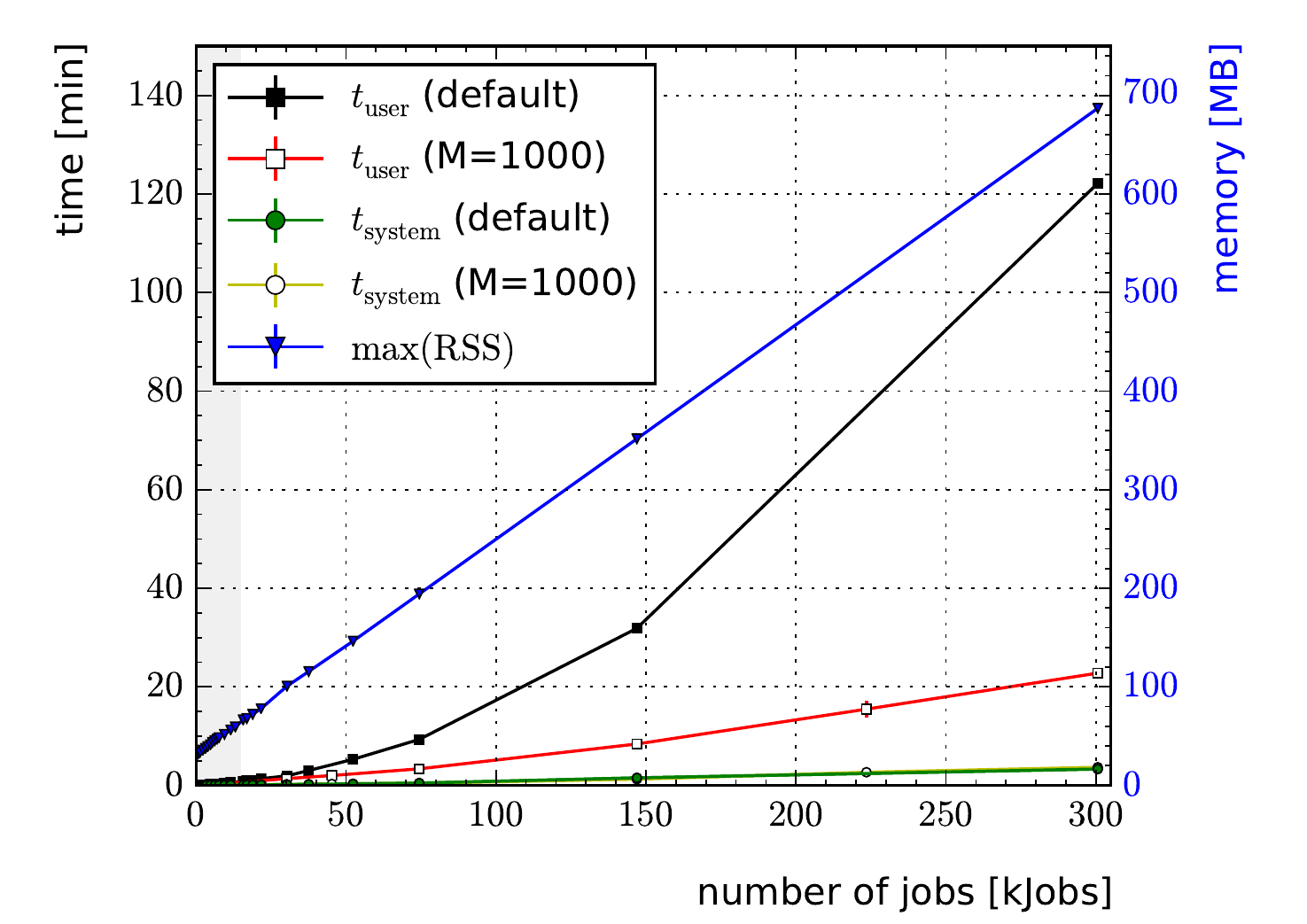}
\caption{\label{SCALING}Memory and CPU time usage for large tasks with up to 300k jobs.}
\end{minipage} 
\end{figure}

The absolute results of these measurements are shown in figure~\ref{SCALING_ZOOM} and~\ref{SCALING}.
The measurements are performed for default settings (filled markers)
and a customized chunk size (open markers) of $M=1000$ that is described in more detail below.

The memory and system time usage scales linearly between 1 and 300k jobs.
In addition to a baseline memory usage of $30$ MB per task,
the used memory scales with $2.3$ KB/job.
This is due to the fact that the default job management plugin keeps all
job management information in memory.
In case the memory consumption should become a problem, a different job management
plugin could access this information from disk,
trading lower memory usage for higher latencies.

\begin{figure}[ht]
\includegraphics[width=0.48\linewidth]{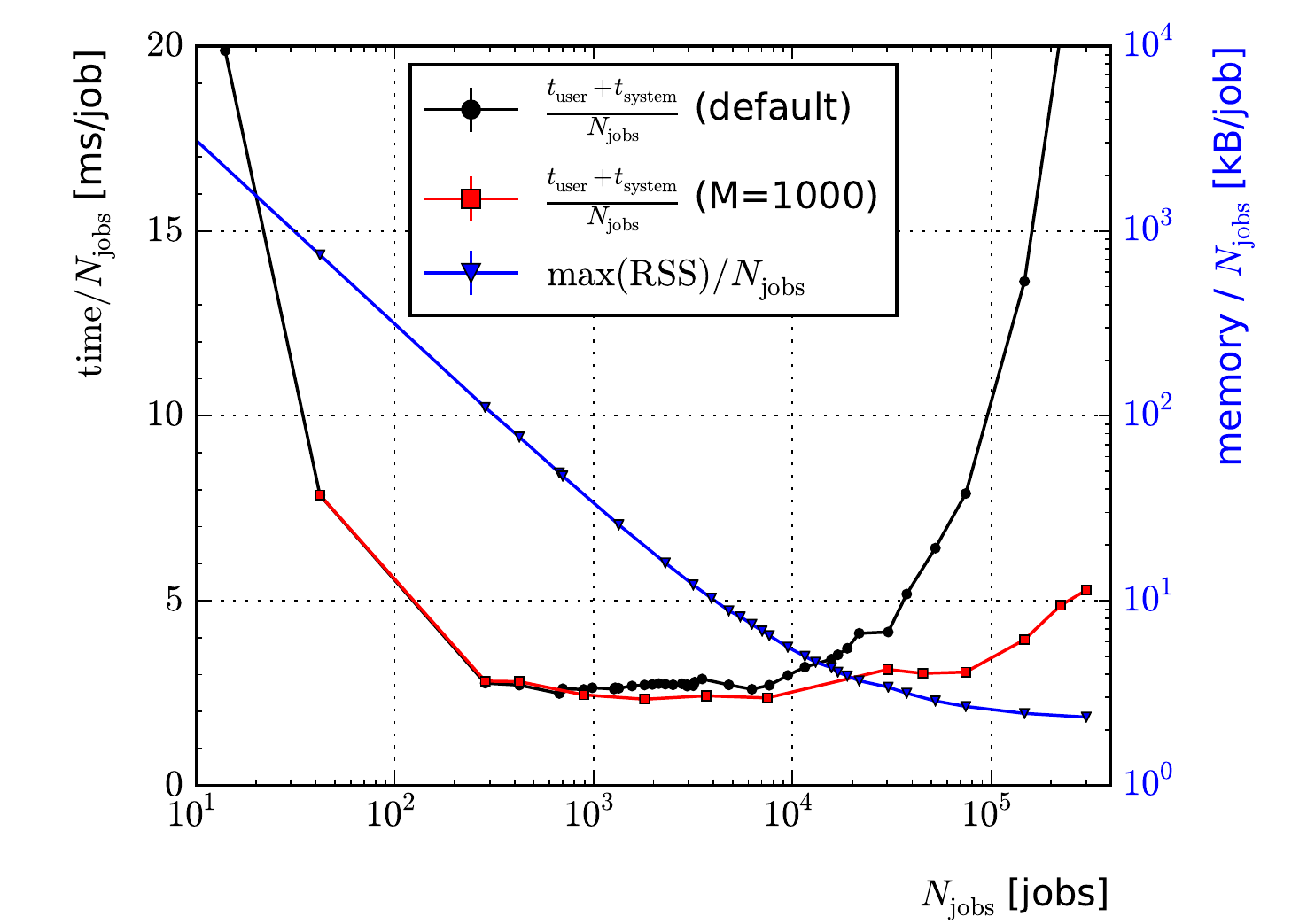}\hspace{0.04\linewidth}%
\begin{minipage}[b]{0.48\linewidth}\caption{\label{SCALING_REL}Memory and total
CPU time consumption relative to the number of jobs.
Measurements with default settings (black markers) and customized chunk size $M=1000$ (red markers) show
differences between the total CPU time per job for large task sizes due to sorting overhead.
The fixed memory overhead of $30$ MB per task dominates the memory usage per job for small
task sizes.}
\end{minipage}
\end{figure}

Figure~\ref{SCALING_REL} shows the performance figures relative to the number of jobs.

With an initial $0.25 \mathrm{s}$ start-up time, the time spent by {\it{grid-control}}
itself on a single job over its whole lifetime is around $2.3 \mathrm{ms}/\mathrm{job}$
for most common task sizes. Deviations from this number can be observed for large task sizes.

By default, the full list of jobs with size $N$, which is selected for submission,
status query or retrieval is sorted before a chunk of size $M$ is processed in one pass.
Therefore, the number of sort operations (with an average complexity of $N\log N$) scales with $N/M$.

In order to give users immediate feedback during processing, the default chunk size
is relatively small with $M = 100$ jobs.
For small tasks with $N < 10$k jobs, the time spent on sorting is negligible,
which leads to a linear relationship between the processing time and $N$.

In order to improve the performance for larger task sizes, the chunk size can be configured.
Figure~\ref{SCALING} and \ref{SCALING_REL} demonstrates the CPU time reduction 
for a chunk size of $M = 1$k jobs.

\section{Development and architecture}

Development of {\it{grid-control}} began more than 10 years ago on February 9th 2007.
A week later on February 16th, the first version with support for submitting
CMSSW grid jobs was released. {\it{grid-control}} has been constantly evolving ever
since and continues to be open to contributions on github.
While it has a comprehensive set of functionality, {\it{grid-control}} is
still a very manageable project with just about 20k lines of code when neglecting 3rd party libraries.

\subsection{Unit tests}

\begin{figure}[ht]
\centering
\includegraphics[width=0.8\linewidth]{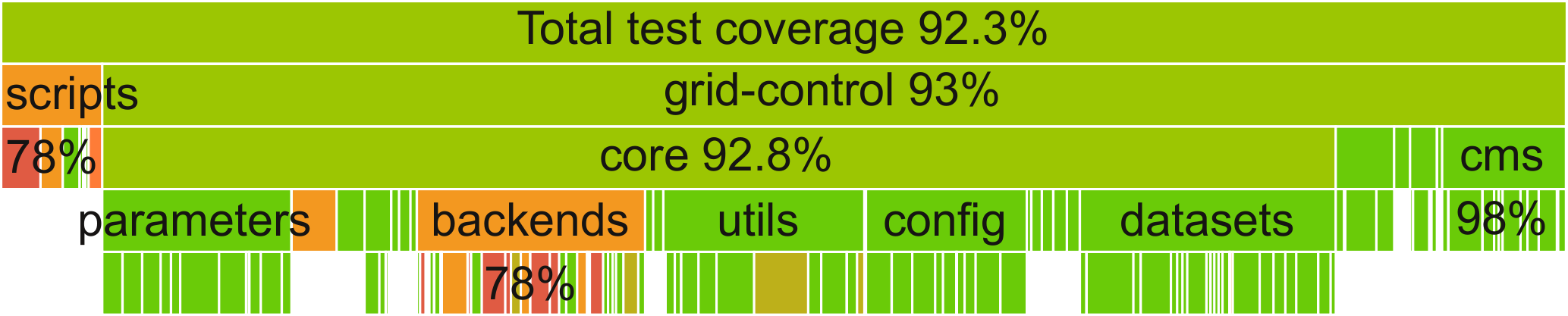}
\caption{\label{COVERAGE}Overview of the test suite coverage.
The widths of the bars are proportional to the size of the source code files / packages
and the colour corresponds to the test suite coverage.
This figure also demonstrates that the CMS experiment
specific plugins only make up a small fraction of the total source code.}
\end{figure}

The source code is automatically tested by the Travis CI service using the default
linux docker container with CPython 2.6-3.7 and pypy/pypy3. Tests with older
versions (Python 2.3-2.5) are performed manually on a regular basis.
The code coverage is reviewed using the codecov.io service, while code
convention and quality checks are done with landscape.io.
Figure~\ref{COVERAGE} shows the code coverage at the beginning of the CHEP 2016 conference.

Achieving a high code coverage is challenging, since code that depends
on external software (e.g. job submission backends) or external services
(e.g. monitoring plugins) need execution environments that simulate
the behaviour (and possible misbehaviour) of these dependencies.

The CMS experiment software for example is currently tested with a mock environment
and a self-written cmsRun script that generates output according to the
given config files. This allows much faster turnaround times compared to using a CMSSW docker image.

Since the CMS data services require authentication
with a private grid proxy, the CMS data provider plugins are tested with a web
service query plugin that responds with pre-recorded answers.
However the REST query plugins are tested against the public httpbin.org service.

\subsection{Documentation}

In addition to a large set of examples that cover most use cases and a
small tutorial, {\it{grid-control}} comes with a comprehensive list of all
available configuration options (currently more than 530) together
with their default value, type, and scope\cite{GCOPTS}. This list is created by
a custom script directly from the Python AST and ensures that no
option in the source code is left undocumented.

\subsection{Architecture}

{\it{grid-control}} is using a highly modular design, where all functionality
is provided by specialized plugins with well-defined interfaces.
The core package provides around 300 plugins, most of which belong
to one of several larger categories of plugins with common interfaces.
The files that contain these plugins are loaded only on demand, which
reduces the start-up time considerably.

An average running workflow is composed of around 100 active plugins, which
take care of all different aspects from parsing job status queries,
handling monitoring events, filtering dataset location lists to writing
the job database.

The interplay between these plugins is configured through a sophisticated
configuration system that allows plugins to configure other plugins (e.g. to provide
context dependent default settings) and which can react to changes to the
user configuration.

During this configuration phase at the start of the program, a large number of
additional short-lived plugins is active, since the composition of the workflow
itself is highly configurable and performed by dedicated configuration plugins.

\section{CMS experiment support}

The CMS experiment software integration (with full support for
CMSSW version 1.x-8.x), is realized with a small number of plugins,
distributed in an optional package that can be safely deleted if
it is not needed.

Two data source plugins allow to either query DAS\cite{DAS} or
DBS\cite{DBS2},\cite{DBS3} together with PhEDEx\cite{PHEDEX}
to retrieve official CMS dataset information. This includes event counts,
block location and other metadata such as run / luminosity section information.
Additional dataset pipeline plugins provide the functionality to filter
and split datasets along run / luminosity section ranges.

Dedicated Task modules take care of setting up and running the
official CMS experiment software and storing metadata that allows
subsequent registration of the output files in DBS.

A monitoring plugin is available that can be used to report
configurable task and job status information to the CMS Dashboard\cite{DASHBOARD}.

In addition to these plugins, the CMS experiment support also include
several scripts to simplify common CMS specific tasks.
These scripts allow to create JSON reports about the processed
run / luminosity sections for luminosity calculations
or the registration of output datasets in DBS to make them available to the collaboration.

The DBS registration script can even handle parameterized jobs that perform
multiple reprocessing steps in the same job
and produce multiple output files. This kind of advanced workflow is common
for example in CMS tau embedding studies and currently not supported by other user tools.
\vspace{-1em}
\section*{\ack}
The authors would like to thank the following people for their
contributions to the development of {\it{grid-control}}:

Fred Stober, Max Fischer, Jana Saout, Manuel Giffels, Thomas Hauth, Armin Scheurer, Andreas Oehler,
Matthias Wolf, Volker B\"uge, Jochen Ott, Thomas M\"uller, Manuel Zeise, Klaus Rabbertz, Armin Burgmeier,
Artur Akhmetshin, Johannes Lange, Frank Fischer, Joosep Pata, Oliver Oberst, Eike Schlieckau,
Benjamin Klein, Joram Berger, Benjamin Treiber, Dominik Haitz, Stefan Wayand, Luigi Calligaris,
Raphael Friese, Peter Krauss, Matthias Schnepf, Georg Sieber, Corinna G\"unth, Philipp Schieferdecker,
Georg Fleig, Hauke Held, Gregor Kasieczka, Jasmin Kiefer and Thomas Scharrer.

Most people worked on {\it{grid-control}} in their spare time in order to enable
or improve the efficiency of some particular workflows that were needed as
part of their LHC data analysis efforts.
The authors acknowledge that the majority of these LHC data analysis efforts
were funded by the German Federal Ministry of Education and Research (BMBF),
the Helmholtz Association (HGF) and the German State Ministry for Research of Baden-W\"urttemberg.

\section*{References}
\bibliography{gc_proc}
\bibliographystyle{iopart-num}

\end{document}